\documentclass[aps,prx,reprint,superscriptaddress,footnote, twocolumn]{revtex4-1}

\usepackage{appendix}
\usepackage[export]{adjustbox}
\usepackage{subcaption}

\usepackage{tikz}  
\usetikzlibrary{arrows,shapes,positioning,shadows,backgrounds,fit}

\usepackage{enumerate}
\usepackage{relsize}
\usepackage{graphicx}
\usepackage{amsmath, bbm}
\usepackage{amssymb}
\usepackage{amsthm}
\usepackage{mathrsfs}
\usepackage{xcolor}
\usepackage{cancel}
\usepackage{titlesec}
\usepackage{enumitem}  
\usepackage{mathtools}
\usepackage{braket}

\usepackage[normalem]{ulem}

\titlespacing{\section}{0ex}{2ex}{0.4ex}
\titleformat{name=\section}{\bf}{\thesection.}{.3em}{}

\usepackage{soul}
\usepackage{dsfont}

\def\be{\begin{eqnarray}}
\def\ee{\end{eqnarray}}
\newcommand{\tr}[1]{\text{Tr}\left(#1\right)}
\newcommand{\Tr}[1]{\text{Tr}\left[#1\right]}

\usepackage{amsthm}

\theoremstyle{plain}

\newtheorem{cor}{Corollary}
\newtheorem{thm}{Theorem}

\definecolor{myblue}{rgb}{0.2,0.2,0.8}
\definecolor{myblack}{rgb}{0,0,0}
\definecolor{myurl}{rgb}{0.1,0.1,0.4}

\usepackage[colorlinks=true,citecolor=myblue,linkcolor=myblack,urlcolor=myurl]{hyperref}

\begin{document}

\title{Entropy-based random quantum states}

\date{\today}

\author{Harry J.~D. Miller}
\affiliation{Department of Physics and Astronomy, The University of Manchester, Manchester M13 9PL, UK}

\begin{abstract}
In quantum information geometry, the curvature of von-Neumann entropy and  relative entropy  induce a natural metric on the space of mixed quantum states. Here we use this information metric to construct a random matrix ensemble for states and investigate its key statistical properties such as the asymptotic eigenvalue density and mean  entropy. We present an algorithm for generating these entropy-based random density matrices, thus providing a new recipe for random state generation that differs from the well established Hilbert-Schmidt and Bures-Hall ensemble approaches. We also prove a duality between the entropy-based state ensemble and a random Hamiltonian model constructed from the thermodynamic length over the set of Gibbs states. This Hamiltonian model is found to display Wigner level repulsion, implying that the dual   state ensemble can be realised as a random Gibbs state with respect to a class of chaotic Hamiltonians. As an application we use our model to compute the survival probability of a randomly evolved thermofield double state, predicting a ramp and plateau over time that is characteristic of quantum chaos. For other applications, the entropy-based ensemble can be used as  an uninformative prior for Bayesian quantum state or Hamiltonian tomography.
\end{abstract}

\maketitle

\section{Introduction}

\

Random matrix theory plays a key role throughout quantum information theory \cite{collins2016random} and is a core theoretical tool for studying many-body chaos \cite{d2016quantum}. One  application is as a mathematical framework for characterising random quantum states in finite dimensional systems \cite{zyczkowski2011generating}, which are represented by normalised, positive density matrices. Random density matrices capture the typical properties of mixed quantum states such as the entanglement statistics of subsystems in high dimensional Hilbert spaces \cite{page1993average,hayden2006aspects,vznidarivc2006entanglement,nadal2010phase,aubrun2014entanglement,vivo2016random,bhattacharjee2021eigenstate,chang2019evolution,bianchi2022volume}. They are also a main ingredient in Bayesian state inference and discrimination tasks, where random state ensembles can be used to construct  uninformative priors \cite{buvzek1998reconstruction,sentis2010multicopy,schmied2016quantum,granade2016practical,lohani2023demonstration,Boeyens2025}. Further applications include models of quantum chaos \cite{kubotani2008exact,demkowicz2004global,sarkar2021generation}, deep thermalisation \cite{sherry2025mixed,yu2025mixed}, black hole physics \cite{page1993information,kudler2021relative}, and many-body work extraction \cite{hovhannisyan2024concentration}.

Statistical measures over random density matrices can be constructed with recourse to the geometry of quantum state space \cite{bengtsson2017geometry}. There are currently two studied geometric-induced ensembles over states known as the Hilbert-Schmidt \cite{zyczkowski2001induced,zyczkowski2003hilbert}  and Bures-Hall measures \cite{hall1998random,slater1999hall,sommers2003bures}. Both ensembles have known procedures for generation and can each be obtained as subsystems of a global random pure state \cite{al2010random,zyczkowski2011generating}. However, there is scope for exploring further types of ensembles from the wide family of possible metrics over density matrices \cite{petz1996monotone}, and it remains an ongoing question to understand which ensembles are most appropriate for characterising randomness in different types of physical system.

In order to expand these investigations of random state ensembles in quantum information theory, we are motivated to consider an alternative information-geometric approach that is fundamentally tied to quantum entropy and hypothesis testing. Balian proposed a natural information-based metric over quantum state space using the curvature of von-Neumann entropy \cite{balian1986dissipation,balian2014entropy}. In the literature this is often known as the Bogoliubov-Kubo-Mori inner product \cite{petz1994geometry} due to its relevance to linear response theory in statistical mechanics \cite{mori1965transport,naudts1975linear}. The metric is physically motivated as it provides a  quantum generalisation of thermodynamic length \cite{crooks2007measuring} and can be used to characterise paths of minimal entropy production in dissipative quantum systems \cite{scandi2019thermodynamic,abiuso2020geometric}. Furthermore, the metric has found a wider range of applications such as the characterisation of phase transitions in critical systems \cite{janyszek1989riemannian}, information-geometric approaches to statistical mechanics \cite{floerchinger2020thermodynamics}, conformal field theory \cite{lashkari2016canonical,czech2023changing,de2023quantum}, optimal transport \cite{carlen2020non}, and quantum estimation theory \cite{hayashi2002two}.  

In this paper we propose to use the entropy-based metric as an uninformative probability measure for random states. Using the fact that the metric arises as a second order expansion of quantum relative entropy \cite{hayashi2002two}, we argue that this measure is operationally motivated in the sense that probability density can be assigned based on the degree of asymptotic distinguishability between regions of state space.  We then demonstrate how to sample quantum states with respect to this entropy-based metric, providing a new method for generating random density matrices. We find the global eigenvalue density and derive an analogue of Page's formula \cite{page1993average} for the average entropy of a high dimensional random state. 

In conjunction with the state ensemble, we also introduce a random Hamiltonian model built from the measure of thermodynamic length \cite{janyszek1989riemannian,scandi2019thermodynamic,mehboudi2022thermodynamic} over the manifold of Gibbs thermal states. This is used to prove that the entropy-based ensemble is equivalent to a distribution of random Gibbs states, thus yielding a clear thermodynamic interpretation of the measure. While globally distinct from the standard GUE ensemble, our thermodynamic Hamiltonian model displays characteristic Wigner level repulsion and can therefore be used to model quantum chaos. We illustrate this by applying the model to the dynamics of thermofield double states and find signatures of quantum chaos in the time evolution of the survival probability. 

The paper is organised as follows: Section~\ref{sec:1} we review the Hilbert-Schmidt and Bures-Hall random matrix ensembles, Section~\ref{sec:2} presents our proposed entropy-based ensemble, in Section~\ref{sec:3} we analyse the asymptotic properties of the ensemble, in Section~\ref{sec:4} we prove a measure concentration result, in Section~\ref{sec:alg} we present an algorithm for sampling the entropy-based random states, Section~\ref{sec:6} introduces the dual Hamiltonian ensemble using the thermodynamic length, and Section~\ref{sec:7} applies our model to the chaotic dynamics of a thermofield double state.

\section{Metric-induced random density matrices: known approaches}\label{sec:1}

\

We begin with an overview of preexisting methods for generating random quantum states. Let $\mathcal{H}\simeq \mathbb{C}^N$ be an $N$-dimensional Hilbert space and $\mathcal{S}(\mathcal{H})$ be the set of positive, normalised and full-rank density matrices. Random pure quantum states are uniquely characterised by the Haar distributed elements of the projective Hilbert space $\mathcal{P}(\mathcal{H})$ \cite{gibbons1992typical,wootters1990random}. When extending this to define random density matrices this uniqueness no longer holds and different distributions may be defined with different statistical properties \cite{zyczkowski2011generating}. There are currently two known canonical recipes for generating a random density matrix from an uninformative prior. For the simplest method one begins by taking a random pure state from the doubled Hilbert space $\ket{\phi}\in \mathcal{H}\otimes \mathcal{\bar{H}}$ and then takes a partial trace to obtain the density matrix
\begin{align}\label{eq:HS}
    \rho_{HS}=\text{Tr}_{\mathcal{\bar{H}}} \ket{\phi}\bra{\phi}=\frac{A A^\dagger}{\tr{A A^\dagger}},
\end{align}
Here $A$ is a complex Gaussian  $N\times N$ Ginibre matrix, and the resulting matrix $\rho_{HS}$ is distributed according to the Hilbert-Schmidt (HS) measure \cite{zyczkowski2003hilbert}. Its statistical features have been extensively investigated particularly in the context of entanglement entropy statistics of typical pure states  \cite{page1993average,sanchez1995simple,sommers2004statistical,link2015geometry,christandl2014eigenvalue,kumar2020wishart}. The eigenvalues of the reduced density matrix $\vec{r}=(r_1,\cdots,r_N)\in\mathbb{R}^N_+$ follow the well-known distribution
\begin{align}\label{eq:pdf}
    P_{HS}(\vec{r})\propto \prod_{\nu<\mu} (r_\mu-r_\nu)^2,
\end{align}
with normalisation constraint $\sum^{N}_{k=1}r_k=1$. An alternative method can be used to reduce the bias towards the more mixed states  \cite{hall1998random}; here one takes two copies of a pure random state, applies a local Haar random unitary $U$ and creates a superposition $\ket{\psi}=\frac{1}{\sqrt{2}} (\ket{\phi}+(U\otimes \mathbb{I}_{\mathcal{\bar{H}}}) \ket{\phi})$ \cite{al2010random}. The reduced density matrix 
\begin{align}\label{eq:BH}
    \rho_{BH}=\text{Tr}_{\mathcal{\bar{H}}} \ket{\psi}\bra{\psi}=\frac{(\mathbb{I}_N+U)A A^\dagger(\mathbb{I}_N+U^\dagger)}{\tr{(\mathbb{I}_N+U)A A^\dagger(\mathbb{I}_N+U^\dagger)}},
\end{align}
is then distributed by the Bures-Hall measure \cite{hall1998random,slater1999hall,sommers2003bures}, and statistical properties of these states have also been studied \cite{zyczkowski2005average,zyczkowski2011generating,borot2012purity,sarkar2019bures,wei2020exact,hovhannisyan2024concentration}. Eigenvalues are distributed according to
\begin{align}\label{eq:pdf}
    P_{BH}(\vec{r})\propto \prod_{k=1}^{N} \frac{1}{\sqrt{r_k}}\prod_{\nu<\mu} \frac{(r_\mu-r_\nu)^2}{r_\mu+r_\nu}.
\end{align}
In comparison to the Hilbert-Schmidt states, the Bures-Hall states are less mixed on average \cite{sarkar2019bures}, since the measure places more weight towards the boundary of $\mathcal{S}(\mathcal{H})$. In contrast to the HS measure, the Bures metric is Fisher-adjusted and can be interpreted as a quantum version of Jeffreys  prior \cite{amari2016information}. It is therefore often employed as an uninformative prior in Bayesian quantum tomography \cite{lohani2021improving}. Taken together, the HS and BH  ensembles are the two canonical choices for describing distributions of random density matrices in quantum theory in the absence of additional information or constraints.

\

\section{The entropy-based metric and its volume form}\label{sec:2}

\

As we argue here, the HS and BH measures are not the only metric-induced ensembles one can construct in quantum information geometry. One alternative possibility is to use an information-based distance between quantum states via the von-Neumann entropy \cite{von2018mathematical},
\begin{align}
    S(\rho)=-\tr{\rho \ \text{ln} \ \rho}.
\end{align}
As a concave function of the state, the negative of its Hessian form gives rise to an inner product on $\mathcal{S}(\mathcal{H})$ and the squared infinitesimal distance between states $\rho$ and $\rho+\delta\rho$ is given by \cite{balian1986dissipation}: 
\begin{align}\label{eq:entmet}
    ds^2:=-d^2 S(\rho)=\sum_{\nu,\mu} \bigg[\frac{\text{ln} \ r_\nu-\text{ln} \ r_\mu}{r_\nu-r_\mu}\bigg] \big| d\rho_{\nu\mu}\big|^2,
\end{align}
which is here expressed in terms of the diagonal decomposition of a state $\rho=\sum_\nu r_\nu \ket{\nu}\bra{\nu}$ and variational elements $d\rho_{\nu\mu}=\bra{\mu}d\rho\ket{\nu}$. This defines the Bogoliubov-Kubo-Mori (BKM) metric over quantum states \cite{petz1994geometry}. Intuitively, we may interpret this distance directly in terms of information loss due to mixing \cite{balian2014entropy}; considering two close states $\rho_1=\rho-\frac{1}{2}\delta\rho$ and $\rho_2=\rho+\frac{1}{2}\delta\rho$, the information lost by taking their convex mixture is directly measured by the squared distance $\frac{1}{8}ds^2 \sim \Delta S=S(\rho)-\frac{1}{2}[S(\rho_1)+S(\rho_2)]$. Alternatively, we may arrive at the same metric from the second-order variations of the quantum relative entropy $S(\rho||\sigma)=\tr{\rho \ (\text{ln} \ \rho-\text{ln} \ \sigma)}$ \cite{lindblad1974expectations}, since
\begin{align}
    ds^2=-\frac{\partial^2 }{\partial t \partial s}S(\rho+t\delta \rho \ || \ \rho+s\delta \rho)\bigg|_{s=t=0}.
\end{align}
The BKM metric belongs to the family of quantum generalisations of the classical Fisher-Rao metric \cite{amari2016information} which can all be characterised by second-order variations of monotone divergence measures between states \cite{petz1996monotone,lesniewski1999monotone,scandi2023quantum}. While the curvature properties of the BKM metric have been well-studied \cite{petz1994geometry,dittmann2000curvature,michor2000curvature,gibilisco2005monotonicity}, little is known about its role as a measure over $\mathcal{S}(\mathcal{H})$.

\begin{figure*}[t]
  \centering
  \begin{subfigure}[t]{0.49\textwidth}
    \centering
    \includegraphics[width=\linewidth]{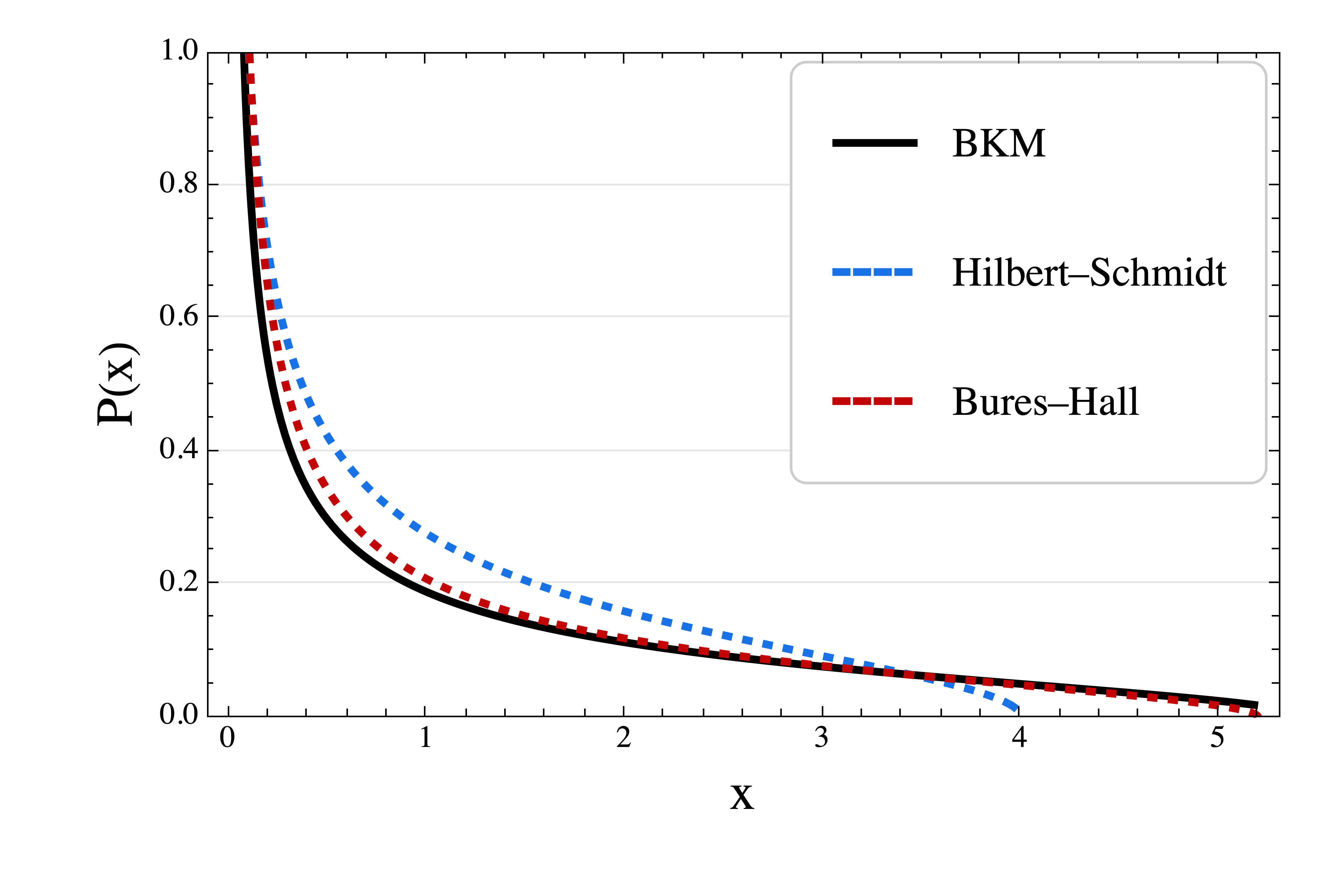}%
    \caption{Asymptotic marginal eigenvalue density}
    \label{fig:lam}
  \end{subfigure}\hfill%
  \begin{subfigure}[t]{0.47\textwidth}
    \centering
    \raisebox{1em}{
    \includegraphics[width=0.94\linewidth]{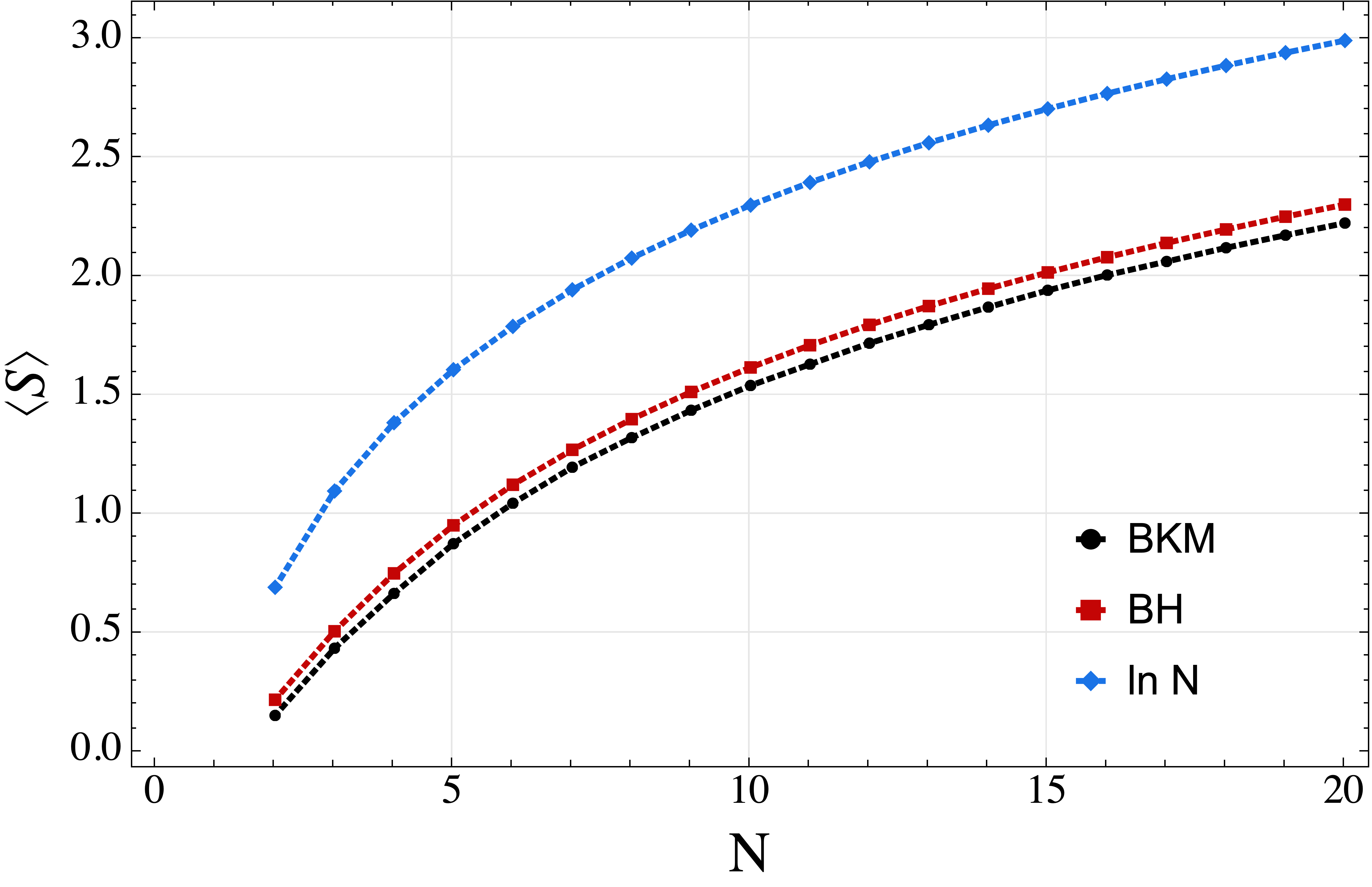}}
    \caption{Average entropy versus dimension}
    \label{fig:mean}
  \end{subfigure}
  \caption{(a) Plot of the asymptotic marginal eigenvalue density for the BKM ensemble~\eqref{eq:lambert}, compared with the equivalent distribution for Hilbert--Schmidt states~\eqref{eq:HS} and Bures--Hall states~\eqref{eq:BH}. (b) Plot of the average von Neumann entropy $\langle S\rangle$ as a function of dimension $N$ for the BKM ensemble (black) and Bures--Hall ensemble (red), along with the maximum entropy $\ln N$ (blue).}
\end{figure*}

Here we explore the relationship between the entropy-based metric and random matrix theory, which will lead us to a new method for sampling random density matrices. The central idea is to build a random measure on $\mathcal{S}(\mathcal{H})$ from the corresponding Riemannian volume form of $ds^2=-d^2 S(\rho)$, denoted $dV(\rho)$. This measure can be motivated operationally by considering the asymptotic distinguishability between neighboring density matrices, following a similar line of reasoning that is used to define Jeffreys' prior in classical inference \cite{ly2017tutorial}. As is well known in quantum information theory \cite{hayashi2017quantum}, $n$ copies of two states $\rho$ and $\rho+d\rho$ are practically indistinguishable via a binary quantum hypothesis test when their relative entropy is sufficiently small as $n\to\infty$, namely 
\begin{align}
    n S(\rho+d\rho|| \rho )\lesssim \mathcal{O}(1), 
\end{align}
For close states one has the local approximation to relative entropy via the BKM metric \cite{hayashi2002two}
\begin{align}
    S(\rho+d\rho|| \rho)=\frac{1}{2}ds^2+\mathcal{O}(d\rho^3),
\end{align}
and so we can therefore define a \textit{indistinguishability ball} $\mathcal{B}_n(\rho)$ around a state $\rho$ defined by
\begin{align}
    \mathcal{B}_n(\rho):=\bigg\{\rho+d\rho \in\mathcal{S}(\mathcal{H}): \ \ ds\leq \sqrt{\frac{2}{n}}\bigg\}
\end{align}
For any given resolution set by $n$, the Lebesgue volume of this ball is proportional to the inverse square root of the metric determinant,
\begin{align}\label{eq:ball}
    \text{Vol}(\mathcal{B}_n(\rho))\propto \frac{1}{\sqrt{\text{det} \ g(\rho)}},
\end{align}
where $g(\rho)$ denotes the metric tensor associated with squared distance $ds^2=g_{ij}(\rho)dx^i dx^j$ and $\vec{x}\in\mathbb{R}^{N^2-1}$  a set of affine real coordinates for the state $\rho=\rho(\vec{x})$. Now we may define a normalised probability measure $P(\rho)$ over density matrices by counting how many of them in this small neighbourhood are asymptotically distinguishable via such a hypothesis test, up to resolution set by $n$. By~\eqref{eq:ball} this probability density must be independent of $n$ and proportional to the coordinate-invariant Riemannian volume form $dV(\rho)=\sqrt{\text{det} \ g(\rho)}d^{N^2-1}x$, thus 
\begin{align}
P(\rho)\mathcal{D}\rho=\frac{dV(\rho)}{\text{Vol}(N)}, \ \ \ \text{Vol}(N)=\int_{\mathcal{S}(\mathcal{H})} dV(\rho)
\end{align}
Here we have defined the probability density with respect to the uniform Lebesgue measure over hermitian matrices,
\begin{align}
\mathcal{D}\rho=\prod^N_{k=1}d\rho_k \prod_{\nu<\mu} d( \Re\rho_{\nu\mu})d( \Im \rho_{\nu\mu})
\end{align}
The volume form itself can be found straightforwardly from diagonalising the metric in~\eqref{eq:entmet}, leading to
\begin{align}\label{eq:vol}
    dV(\rho)=\text{det}(\rho)^{-1/2}\frac{\Delta(\text{ln} \ \rho)}{\Delta(\rho)}\Theta(\rho)\delta(1-\tr{\rho})\mathcal{D}\rho.
\end{align}
where $\Delta(\rho)=\prod_{\nu<\mu}(r_\mu-r_\nu)$ is the Vandermonde determinant of the state $\rho=\sum_k r_k \ket{k}\bra{k}$, and $\Theta(\rho)$ restricts the measure to positive matrices. The probability density is therefore
\begin{align}\label{eq:measure}
    P(\rho)=\frac{1}{\text{Vol}(N)}\frac{\Delta(\text{ln} \ \rho)}{ \text{det}(\sqrt{\rho})\Delta(\rho)}\Theta(\rho)\delta(1-\tr{\rho}).
\end{align}
This is the central theoretical construction of this paper. Much like the Hilbert-Schmidt and Bures-Hall distributions usually employed, we can calculate the statistical properties of randomly sampled states according to measure~\eqref{eq:measure}, and we will refer to it as the BKM ensemble. We note that not all monotone metrics are normalisable \cite{andai2006volume}, though we will prove affirmatively that the BKM ensemble is a valid probability density on the state space. Of particular interest is the marginal distribution of the state eigenvalues. To derive it we plug in the diagonal decomposition $\rho=U \text{diag}(r_1,\cdots r_N)U^\dagger$ into~\eqref{eq:vol} in which case the measure factorises as $P(\rho)\mathcal{D}\rho= P(\vec{r})d^N\vec{r}\times d\mu_{U}(U)$ where $d\mu_{U}(U)$ denotes the Haar unitary measure and $P(\vec{r})$ is a distribution over the positive-definite eigenvalues given by:
\begin{align}\label{eq:pdf}
    \boxed{ \quad P(\vec{r})=C_N \  \text{det}[\rho^{-1/2}] \Delta(\rho)\Delta(\text{ln} \ \rho) \quad},
\end{align}
subject to the normalisation constraint $\sum^{N}_{k=1}r_k=1$, and $C_N$ is a normalisation constant. As we show in Appendix~\ref{app:A}, the constant is
\begin{align}\label{eq:const}
    C_N=\frac{\Gamma(N^2/2)}{\pi^{N/2}  \prod^{N}_{k=1} \Gamma(k+1)},
\end{align}
As a normalised probability distribution, moments of unitarily invariant quantities such as entropy and purity can be computed directly from the eigenvalue density~\eqref{eq:pdf}. Comparing this density with the the HS distribution~\eqref{eq:HS} and BH distribution~\eqref{eq:BH}, we see its distinguishing feature is the logarithmic interaction term. Due to the product form of the metric, the total volume of the BKM ensemble can be found by multiplying this with the volume of the Haar measure and dividing by $N!$ to account for the number of permutations of eigenvalues. Adopting the same normalisation convention as \cite{sommers2003bures}, the total volume is proportional to the volume of the flag manifold $\text{Vol}(Fl_N^{\mathbb{C}})$ and we find 
\begin{align}\label{eq:volume}
    \text{Vol}(N)=\frac{\text{Vol}(Fl_N^{\mathbb{C}})}{C_N N!}=2^{N(N-1)/2}\frac{\pi^{N^2/2}}{\Gamma(N^2/2)}
\end{align}
In comparison to the Bures-Hall volume $\text{Vol}_{BH}(N)$ derived in \cite{sommers2003bures}, we find that the BKM volume shrinks significantly slower with increasing $N$ \footnote{Note the convention in \cite{sommers2003bures} is to multiply the line element $ds$ by a factor of $2$, and as an $N^2-1$ dimensional manifold the volume is thus multiplied by a factor of $2^{N^2-1}$.}, since 
\begin{align}
    \text{Vol}(N)=2^{N(N-1)/2}\text{Vol}_{BH}(N).
\end{align}
This ratio between the BH and BKM volumes proves a numerical conjecture of Slater \cite{slater2005silver}.

\section{Asymptotic properties and relation to the Muttalib-Borodin ensemble} \label{sec:3}

\

There is an insightful connection between the measure~\eqref{eq:pdf} and the class of Muttalib--Borodin ensembles in random matrix theory \cite{forrester2017muttalib,forrester2018selberg}. The Muttalib-Borodin ensemble was originally introduced to describe the eigenvalue densities of transition matrices in disordered  quantum transport \cite{muttalib1995random}.  These are positive matrices with a Laguerre-weighted normalised eigenvalue density on $\vec{r}\in\mathbb{R}^N_+$ of form
\begin{align}\label{eq:MB}
    \nonumber &P_{\theta,\alpha}(\vec{r})=C_{\theta,\alpha} \prod^N_{k=1}r_k^\alpha e^{- r_k} \prod_{\nu<\mu} (r_\mu-r_\nu)\frac{(r_\mu^\theta-r_\nu^\theta)}{\theta}, \\
    &C_{\theta,\alpha}=\frac{1}{\prod_{k=1}^N \Gamma(k+1)\Gamma(\theta(k-1)+\alpha+1)},
\end{align}
where $\alpha>-1$ and $\theta>0$. Observe that one can recover the density matrix ensemble~\eqref{eq:pdf} as a fixed-trace version of of the Muttalib--Borodin ensemble by choosing $\alpha=-1/2$ and $\theta\to 0$, using  
\begin{align}
    \lim_{\theta\to 0^+} \frac{(x^\theta-y^\theta)}{\theta}=\text{ln}(x/y).
\end{align}
The fixed trace constraint arises simply by applying a Laplace transform $\mathcal{L}_s[.]$ to the rescaled distribution $P_{\theta,\alpha}(s\vec{r})$, leading to the direct correspondence to the BKM joint density~\eqref{eq:pdf},
\begin{align}
P(\vec{r})=\Gamma(N^2/2)\mathcal{L}^{-1}_s\big[s^{\frac{N^2}{2}-1} P_{0,-1/2}(s\vec{r})\big](t)\bigg|_{t=1} 
\end{align}
For example we can use this correspondence to immediately determine the normalisation constant $C_N$ from $C_{\theta,\alpha}$, in agreement with the calculated value~\eqref{eq:const}.

This connection to the Muttalib--Borodin ensemble is useful because one can extrapolate properties of the BKM ensemble  from known properties of~\eqref{eq:MB}. For example, consider the marginal probability $P(r)$ of a single eigenvalue, obtained from
\begin{align}
    P(r)=\int_0^1 dr_2 \cdots \int_0^1 dr_N \ \delta\bigg(1-\sum^N_{k=1}r_k\bigg)P(\vec{r})\bigg|_{r_1=r}
\end{align}
We can determine the distribution of a rescaled eigenvalue $x=Nr$ in the asymptotic limit $N\to \infty$. Following a general argument that can be applied to log-gases with linear potentials \cite{forrester2010log}, in the asymptotic regime ensemble equivalence ensures that the marginal of the  `microcanonical' measure~\eqref{eq:pdf}  will be equivalent to the marginal of the `canonical' measure~\eqref{eq:MB} (taking again the parameter choices $\theta=0$ and $\alpha=-1/2$) subject to the trace constraint $\int dx \ x P(x)=1$. Then by combining Prop 4.6 of \cite{forrester2017muttalib} with the linear trace constraint, the asymptotic probability of a single rescaled eigenvalue of the BKM ensemble is 
\begin{align}\label{eq:lambert}
    P(x)\underset{N\to\infty}{=} -\Im \bigg[\frac{1}{\pi x W(-2x^{-1})}\bigg], \ \ \ \ x\in[0,2e]
\end{align}
where $W(x)$ is the Lambert $W$-function (ie. the inverse of the map $w\mapsto w e^w$). In Figure~\ref{fig:lam} we plot this function alongside corresponding asymptotic densities for both the HS random states~\eqref{eq:HS} and BH states~\eqref{eq:BH}, both of which can be found in \cite{zyczkowski2011generating}. The density $P(x)$ diverges as $\sim 1/x (\text{ln} \ 2/x)^2$, which is faster than both HS ($\sim 1/\sqrt{x}$) and BH ($\sim x^{-2/3}$) densities. This indicates that the BKM random states are comparatively less mixed, which we can  verify from computing the average entropy. To do so we note that the Mellin transform of $P(x)$ is given by $\mathcal{M}[P](s)=2^{s-1}(s-1)^{s-1}/s\Gamma(s)$ \cite{cheliotis2018triangular}. Then we use the derivative identity 
\begin{align}\label{eq:mellin}
    \int^\infty_0 dx \ P(x)(x \ \text{ln} \ x)=\frac{d}{ds}\mathcal{M}[P](s)\bigg|_{s=2}=\gamma+\text{ln} \ 2-\frac{1}{2}
\end{align}
where $\gamma\simeq 0.577$ is the Euler–Mascheroni constant. Now since the entropy is unitarily invariant and given by $S(\rho)=-\sum^N_{k=1}r_k \text{ln} \ r_k$ it can be computed in the asymptotic limit using the marginal distribution $P(r)$. Rescaling $r=x/N$ and using~\eqref{eq:mellin} we find the average entropy of a random state in the asymptotic regime:
\begin{align}\label{eq:asym}
    \nonumber \langle S(\rho) \rangle&\simeq-N\int^1_0 dr \ P(r)(r \ \text{ln} \ r) \\
    &= \text{ln} \ N-\gamma-\text{ln} \ 2+\frac{1}{2}+\mathcal{O}\bigg(\frac{\text{ln} \ N}{N}\bigg).
\end{align}
In comparison, the average Hilbert-Schmidt entropy follows from Page's formula \cite{page1993average}, $\langle S(\rho) \rangle_{HS}\simeq \text{ln} \ N-1/2 $ for large $N$, while the average Bures-Hall entropy is known to be $\langle S(\rho) \rangle_{BH}\simeq \text{ln} \ N-\text{ln} \ 2 $ \cite{sommers2004statistical}. Therefore the BKM random states are the least mixed overall, since 
\begin{align}\label{eq:ineq}
    \langle S(\rho) \rangle< \langle S(\rho) \rangle_{BH}< \langle S(\rho) \rangle_{HS}.
\end{align}
Another distinguishing feature of the BKM distribution is that it places a greater weighting closer to the boundary of full-rank states. To see this, consider the decreasing function
\begin{align}
    f(u):=\{N\langle r_k\rangle;\ u=k/N\}, \qquad u\in(0,1],
\end{align}
where the eigenvalues are labelled in \emph{decreasing} order. In the large-\(N\) limit, \(f(u)\) is the (deterministic) quantile function associated with the limiting density \(P(x)\) of the rescaled eigenvalues \(x=Nr\). Since the fraction of eigenvalues \emph{larger} than \(\tau\) is asymptotically \(\int_{\tau}^{2e}P(x)\,dx\), we have the relation $u=\int_{f(u)}^{2e} P(x)\,dx$. Differentiating this yields the standard quantile identity $f'(u)=-\frac{1}{P(f(u))}$. For the minimum eigenvalue we require the behaviour as $u\uparrow 1$, where $f(u)\downarrow 0$. Using the small-\(x\) asymptotic $P(x)\simeq [x(\ln(2/x))^2]^{-1}$ for $x\ll 1$, one gets the approximation
\begin{align}
    f'(u)\simeq - f(u)\big(\ln\!\tfrac{2}{f(u)}\big)^2,
\end{align}
which integrates to
\begin{align}\label{eq:f_asym}
    f(u)\simeq 2\,\exp\!\Big(-\frac{1}{1-u}\Big)
    \quad (u\uparrow 1).
\end{align}
Since the minimum corresponds to the extreme right edge \(u\approx 1\), a typical estimate is obtained by taking \(u=1-\frac{1}{N}\), giving
\begin{align}\label{eq:rankbound}
    \langle r_{\min}\rangle\sim \frac{f(1-1/N)}{N}\sim \frac{2 e^{-N}}{N}.
\end{align}
This exponentially small scaling with system size starkly contrasts with the polynomial decay exhibited by both the HS and BH measures \cite{vznidarivc2006entanglement}, and highlights the increased concentration of the measure towards the boundary of full rank density matrices. 

\section{Concentration of measure around the maximally mixed state}\label{sec:4}

\

In addition to the spectral properties of the BKM random states we can also look how the matrix elements $\rho_{ij}$ are distributed. It is a standard fact that any unitarily invariant state ensemble will be centered on the maximally mixed state, such that 
\begin{align}\label{eq:maxmix}
    \langle \rho_{ij} \rangle=\frac{\delta_{ij}}{N}.
\end{align}
Furthermore, we can also confirm that the density matrices also concentrate around the maximally mixed state. In order to find the variance in matrix elements we can use the fact that the BKM ensemble has a product form, $P(\rho)\mathcal{D}\rho=P(\vec{r})d^N\vec{r}\times d\mu_U(U)$. Then expanding $\rho=\sum_{i,j} \rho_{ij} \ket{i}\bra{j}$ in an arbitrary basis we have
\begin{align}
    \rho_{ij}\rho_{kl}=\sum_{a,b}r_a r_b U_{ia}U^*_{ja}U_{kb} U_{lb}^*,
\end{align}
Now evaluate the (normalised) Haar integral first (Weingarten calculus) \cite{mele2024introduction},
\begin{align}\label{eq:haar4}
\nonumber\int d\mu_U(U)\, U_{ia}U^*_{ja}U_{kb}U^*_{lb}
&=\frac{\delta_{ij}\delta_{kl}+\delta_{il}\delta_{kj}\delta_{ab}}{N^2-1} \\
& \ \ \ \ \ \  -\frac{\delta_{ij}\delta_{kl}\delta_{ab}+\delta_{il}\delta_{kj}}{N(N^2-1)} .
\end{align}
Using \(\rho_{ij}=\sum_a r_a\,U_{ia}U^*_{ja}\) and \(\bar{P}=\sum_a \langle r_a^2\rangle=\langle\tr\rho^2\rangle\), this gives
\begin{align}\label{eq:rho2corr}
\langle \rho_{ij}\rho_{kl}\rangle
&=\frac{N-\bar{P}}{N^3-N}\,\delta_{ik}\delta_{jl}
+\frac{N\bar{P}-1}{N^3-N}\,\delta_{il}\delta_{jk}.
\end{align}
where we have defined the mean purity
\begin{align}
    \bar{P}=\langle \tr{\rho^2} \rangle.
\end{align}
Combining this with~\eqref{eq:maxmix} we can express the variance of matrix elements in terms of the average purity,
\begin{align}\label{eq:var}
    \langle |\delta\rho_{ij}|^2 \rangle=\begin{cases}
        \frac{(1+\bar{P})(N-1)}{N^3-N}-\frac{1}{N^2}, \ \ \ \ i=j \\
        \frac{N\bar{P}-1}{N^3-N}, \ \ \ \ \ \ \ \ \ \ \ \ \ \ \ \ \ \ \  i\neq j
    \end{cases}
\end{align}
where $\delta\rho=\rho-\langle \rho \rangle$ denotes  the variation in the state relative to its mean.The remaining step is now to compute the average purity in the asymptotic limit, which can be done using the distribution~\eqref{eq:lambert}. As we previously noted, the Mellin transform of the eigenvalue density $P(x)$ is 
\begin{align}\label{eq:mellin}
    \mathcal{M}[P](s)=\int^\infty_0 dx \ P(x) x^{s-1}=\frac{2^{s-1}(s-1)^{s-1}}{s\Gamma(s)}
\end{align}
The average purity for $N\gg 1$ is then simply found by choosing $s=3$, so that
\begin{align}\label{eq:purity}
    \bar{P}\simeq N \int^1_0 dr \ P(r) r^2= \frac{1}{N}\int^\infty_0 dx \ P(x) x^2=\frac{8}{3N}
\end{align}
Finally plugging this into~\eqref{eq:var} and taking $N\gg 1$ gives us the asymptotic variance
\begin{align}
        \langle |\delta\rho_{ij}|^2 \rangle\sim \frac{5}{3N^3}.
\end{align}
Much like the HS and BH ensembles the deviation from the maximally mixed state goes to zero with rate  $\sim 1/N^{3/2}$.

\

\

\section{Algorithm for generating BKM states and Hamiltonians}\label{sec:alg}

\begin{figure}[!t]
  \centering
  \adjincludegraphics[
    width=\columnwidth,
    trim={0 {0} {.2\width} {0}}, 
    clip                                
  ]{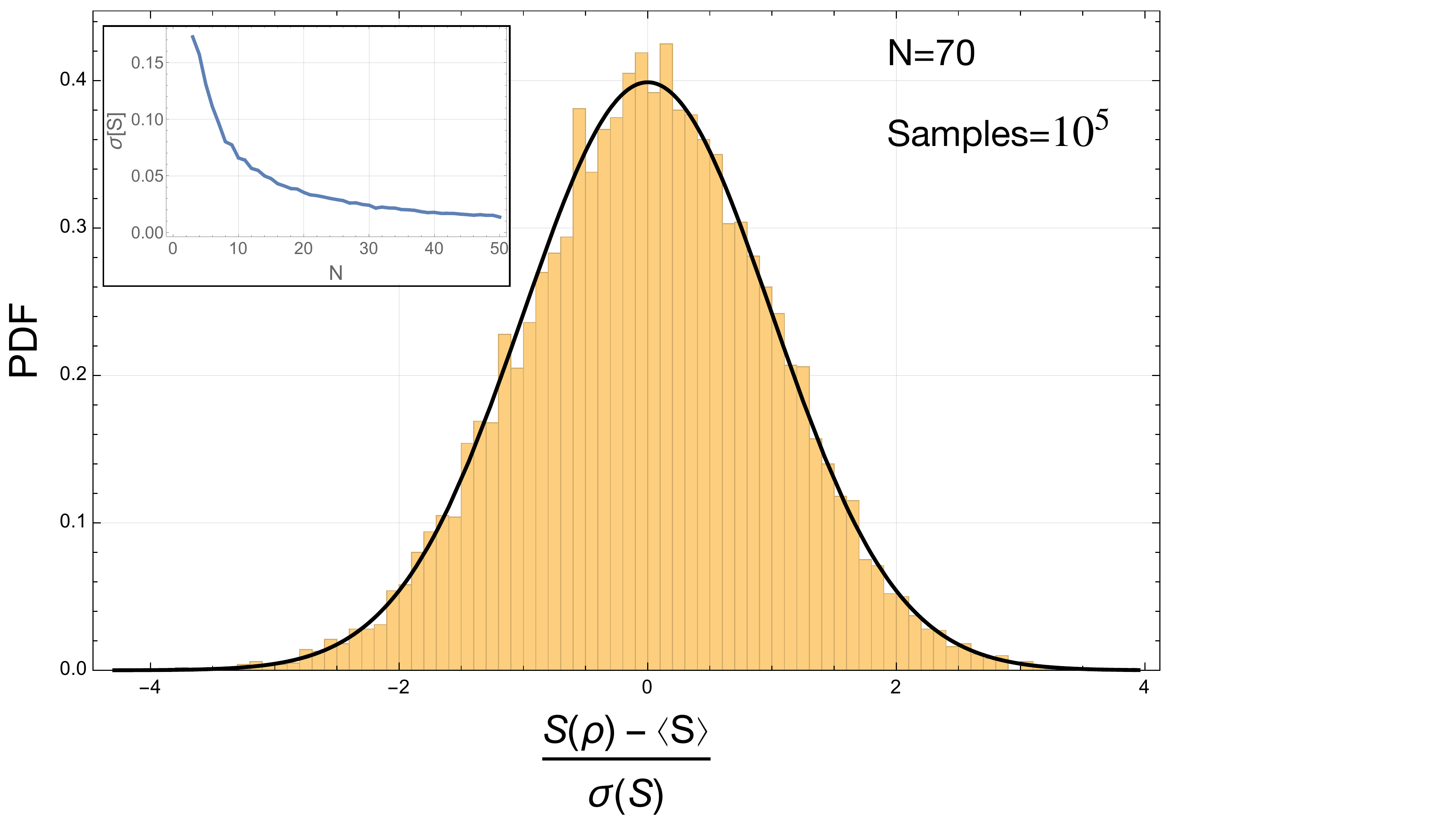}
  \caption{Probability density function of standardised von-Neumann entropy compared with a Gaussian fitting for dimension $N=70$ and $10^5$ samples. \textbf{Inset}: standard deviation in entropy $\sigma[S]$ as a function of $N$.  }
  \label{fig:hist}
\end{figure}

\

It has been shown that the Muttalib-Borodin ensemble~\eqref{eq:MB} can be generated from the singular values of random lower-triangular matrices \cite{cheliotis2018triangular}. Given the correspondence we have established with the BKM distribution we can adapt this recipe to sample the desired ensemble of normalised density matrices. The procedure is summarised in the following theorem:
\begin{thm}\label{thm:algo}
Consider the following matrix sampling procedure
\begin{enumerate}[label=\Alph*)]
    \item Sample a random lower-triangular complex matrix $X=\{X_{jk}\}_{j,k=1}^N$ with non-zero off-diagonal elements $\{X_{jk}: \ 1\leq k<j\leq N\}$ that are complex standard Gaussian variables. The diagonal elements are expressed as
\begin{align}
    X_{kk}=\frac{1}{\sqrt{2}}e^{i\phi_k}Y_k, \ \ k=1,\cdots ,N
\end{align}
where $Y_k$ is sampled from the $\chi_1$-distribution and the phase $\phi_k\in[0,2\pi)$ is uniformly chosen independent of $Y_k$. 
\item Sample a Haar random unitary $U\in U(N)$.
\item Form the normalised density matrix
\begin{align}\label{eq:recipe}
    \rho=\frac{U XX^\dagger U^\dagger}{\Tr{ XX^\dagger }},
\end{align}
\end{enumerate}
The resulting matrix distribution $P(\rho)$ is statistically equivalent to the BKM ensemble~\eqref{eq:measure}.
\end{thm}
\begin{proof}
A full proof is in Appendix~\ref{app:B}. The proof relies on Theorem 2 of \cite{cheliotis2018triangular} which provides the distribution of singular values for the lower-triangular matrices $X$. 
\end{proof}

This  algorithm can be  used to investigate the von-Neumann entropy statistics at finite size. In Figure~\ref{fig:mean} we plot the average entropy $\langle S \rangle$ as a function of dimension $N$ and compare to the maximum entropy $\text{ln} \ N$. Even at moderate size $N\sim 20$ we find that the asymptotic formula~\eqref{eq:asym} is accurate. We also compare this to the average BH entropy, $\langle S \rangle_{BH}=\psi_0(1+N^2/2)-\psi_0(N+1/2)$, derived in \cite{sarkar2019bures,wei2020proof}, where $\psi_0(x)$ is the digamma function. We see that, as anticipated, the BKM states are less mixed overall which confirms the inequality~\eqref{eq:ineq} for finite $N$. In Figure~\ref{fig:hist} we plot a histogram representing the full PDF of entropy in terms of the standardised variable $x=\frac{S(\rho)-\langle S \rangle}{\sigma[S]}$ with $\sigma^2[S]=\langle S^2 \rangle-\langle S \rangle^2$ the variance in entropy. One can see that the PDF is approximately Gaussian, and the inset shows that $\sigma[S]$ decreases monotonically with size $N$. These results give numerical evidence that for BKM states the von-Neumann entropy concentrates on its average~\eqref{eq:asym} at large dimensions. Concentration of measure is also found in both the HS and BH ensembles \cite{hayden2006aspects,bianchi2022volume,wei2020exact,hovhannisyan2024concentration}, and so the BKM ensemble inherits similar concentration properties albeit at less mixedness. A rigorous proof of this concentration of measure is left as an open conjecture.

\section{Thermodynamic length and random Gibbs states}\label{sec:6}

\

In addition to the information-theoretic interpretation of the BKM metric, there is also a fundamental connection to thermodynamics. In this section we will use this correspondence with thermodynamics to provide a direct physical interpretation of the BKM random matrix ensemble. Intuitively, we prove that the BKM density matrices are in fact equilibrium Gibbs states of an underlying random Hamiltonian model. 

The relation between the BKM metric and thermodynamics is best seen by turning attention to the manifold of traceless hermitian matrices,
\begin{align}
    \mathcal{L}_0(\mathcal{H})=\{H=H^\dagger; \ \ \tr{H}=0\}
\end{align}
which has the same dimension as the manifold of density matrices. Now introduce the invertible diffeomorphism 
\begin{align}
    \Phi:\mathcal{L}_0(\mathcal{H})\mapsto \mathcal{S}(\mathcal{H}); \ \ \ \ \ \Phi(H)=\frac{e^{-H}}{\Tr{e^{-H}}},
\end{align}
which defines a Gibbs thermal state $\rho=\Phi(H)$ from any adimensional Hamiltonian $H$. The \textit{free entropy} of a Gibbs state is given by the functional
\begin{align}
    F(H)=\text{ln} \ \Tr{e^{-H}},
\end{align}
which is the Legendre transform of the von-Neumann entropy of the state $\rho=\Phi(H)$. This Legendre duality between $S$ and $F$ carries through to the Riemannian geometry \cite{balian2014entropy}; here we may define the metric on the space of observables $\mathcal{L}_0(\mathcal{H})$ by the Hessian of the free entropy,
\begin{align}
    ds^2=d^2 F(H).
\end{align}
which is the Legendre dual of the BKM metric~\eqref{eq:entmet} for the manifold of states $\mathcal{S}(\mathcal{H})$. This metric defines the \textit{thermodynamic length} \cite{salamon1983thermodynamic,janyszek1989riemannian} on an equilibrium manifold parameterised by the space of Hamiltonians. Under restricted sets of control variables, thermodynamic length is used in quantum thermodynamics to find paths of minimal entropy production \cite{abiuso2020geometric,abiuso2020optimal,mehboudi2022thermodynamic,rolandi2023collective}. Its classical counterpart has a long history of use in the study of fluctuation theory \cite{ruppeiner1995riemannian}, phase transitions \cite{brody2008information} and more recently optimal control in stochastic thermodynamics \cite{crooks2007measuring,machta2015dissipation}. 

Similar to what we did with the BKM ensemble of states, we are now motivated to consider using the thermodynamic length as a metric-induced distribution of random Hamiltonians. This is done by defining a matrix distribution from the corresponding volume form of $ds^2=d^2 F(H)$. By diagonalising the metric in Appendix~\ref{app:C}, this distribution is found to be
\begin{align}\label{eq:pH} 
    P(H)=\frac{\sqrt{N}e^{-\frac{N^2}{2} F(H)}}{\text{Vol}(N)} \bigg|\frac{\Delta(2 e^{-H})}{\Delta(H)}\bigg|\delta(\tr{H}),
\end{align}
which we refer to as the BKM \textit{ Hamiltonian} ensemble. In terms of the energy spectrum $\vec{E}=\{E_1,\cdots, E_N\}$ of the Hamiltonian $H=\sum_k E_k \ket{k}\bra{k}$, we have 
\begin{align}\label{eq:pE} 
    \boxed{\quad P(\vec{E})=\sqrt{N}C_N e^{-\frac{N^2}{2} F(H)} \big|\Delta(2 e^{-H}) \Delta(H)\big|\quad}
\end{align}
with constraint $\sum^N_{k=1}E_k=0$. This energy distribution $P(\vec{E})$ is our second main identification of a random matrix ensemble derived from the entropy-based metric and thermodynamic length, now applied to the set of observables rather than density matrices. The introduction of the  ensemble $P(H)$ provides a direct thermodynamic interpretation underlying the BKM random states distributed by $P(\rho)$. The following theorem proves that they are the set of thermal states obtained from this random Hamiltonian model:
\begin{thm}\label{thm:dual}
The BKM ensemble $P(\rho)$ for density matrices is related to the BKM  Hamiltonian ensemble  $P(H)$ according to
\begin{align}
P(\rho)=\int_{\mathcal{L}_0(\mathcal{H})} \mathcal{D}H \ \delta\bigg[\rho-\frac{e^{-H}}{\textnormal{Tr}(e^{-H})}\bigg]P(H),
\end{align}
\end{thm}
\begin{proof}
The BKM inner product $g_\rho[A,B]$ on $\mathcal{S}(\mathcal{H})$ has the basis independent form \cite{petz1994geometry}
    \begin{align}
\nonumber g_\rho[A,B]&=-\frac{\partial^2}{\partial s \partial t}S(\rho+s A+tB)\bigg|_{s=t=0}, \\
\nonumber&=\tr{B \ \mathbb{J}_\rho^{-1}(A)}, 
\end{align}
where we define the linear operator
\begin{align}
    \nonumber\mathbb{J}_\rho^{-1}(.)=\int^\infty_0 d\lambda \ [\rho+\lambda \mathbb{I} ]^{-1}(.)[\rho+\lambda\mathbb{I} ]^{-1}
\end{align}
On the other hand the BKM inner product $\bar{g}_H[A,B]$ on observables $\mathcal{L}_0(\mathcal{H})$ is
\begin{align}
    \nonumber \bar{g}_{H}[A,B]&=\frac{\partial^2}{\partial s \partial t}F(H+s A+tB)\bigg|_{s=t=0}, \\
    \nonumber&=\tr{B\mathbb{J}_{\Phi(H)}(A)}-\tr{B\Phi(H)}\tr{A\Phi(H)}
\end{align}
We use the identity for the covariant derivative
\begin{align}
    \nonumber d_H\Phi(A)&=\frac{d}{ds}\Phi(H+sA)\bigg|_{s=0}, \\
    \nonumber&=-\int^1_0 ds \ \Phi^s(H)[A-\tr{\Phi(H)A}] \Phi^{1-s}(H)
\end{align}
and combine this with the fact that $\mathbb{J}_{\Phi(H)}\mathbb{J}^{-1}_{\Phi(H)}(A)=A$ to get the relation between the two metrics:
\begin{align}
    \nonumber \bar{g}_H[A,B]=g_{\Phi(H)}[d\Phi_H(A),d\Phi_H(B)]
\end{align}
Thus the diffeomorphism $\Phi$ is an isometry between the Riemannian manifolds $(\mathcal{S}(\mathcal{H}),g)$ and $(\mathcal{L}_0(\mathcal{H}),\bar{g})$. It then follows that the pushforward of the BKM ensemble~\eqref{eq:measure} on the space of states is equal to the BKM ensemble on space of observables~\eqref{eq:pH}, which means for any test function $f(x)$ we have the equality
\begin{align}\label{eq:testf}
    \nonumber \int_{\mathcal{S}(\mathcal{H})}\mathcal{D}\rho \ f(\rho)P(\rho)=\int_{\mathcal{L}_0(\mathcal{H})}\mathcal{D}H \ f(\Phi(H))P(H)
\end{align}
Finally, inserting the integral identity
\begin{align}
    \nonumber f(\Phi(H))=\int_{\mathcal{S}(\mathcal{H})} \mathcal{D}\rho \ f(\rho)\delta[\rho-\Phi(H)],
\end{align}
into the above completes the proof.
\end{proof}
By combining this duality with the algorithm for generating states in Theorem~\ref{thm:algo}, we also have an algorithm for generating random BKM Hamiltonians:  
\begin{cor}\label{cor:alg}
    Consider the following matrix sampling procedure
    \begin{enumerate}[label=\Alph*)]
    \item Sample a random lower-triangular complex matrix $X=\{X_{jk}\}_{j,k=1}^N$ according to Step A of Theorem~\ref{thm:algo}.
    \item Sample a Haar random unitary $U\in U(N)$.
\item Form the traceless hermitian matrix
\begin{align}
    H=-\textnormal{\text{ln}} \ U X X^\dagger U^\dagger+\frac{1}{N}\textnormal{Tr}\big(\textnormal{\text{ln}} \ X X^\dagger\big)\mathbb{I}
\end{align}
    \end{enumerate}
 The resulting matrix distribution $P(H)$ is statistically equivalent to the BKM Hamiltonian ensemble~\eqref{eq:pH}.   
\end{cor}
\begin{proof}
    Follows immediately from the dual relation between $P(\rho)$ and $P(H)$ in Theorem~\ref{thm:dual}.
\end{proof}

\begin{figure}[!t]
  \centering
  \includegraphics[width=\columnwidth]{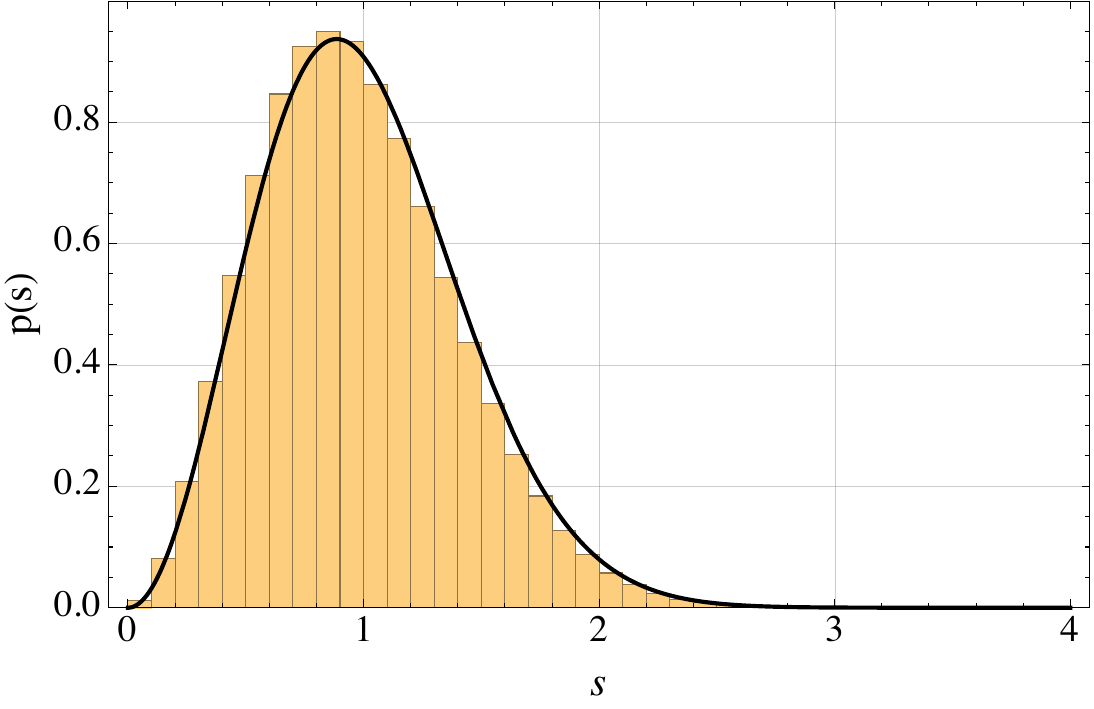}
  \caption{Normalised unfolded gap distribution for the BKM Hamiltonian ensemble~\eqref{eq:pE}, compared with the GUE Wigner surmise~\eqref{eq:wigner} (solid line). System size is $N=100$ and number of samples is $10^4$.  }
  \label{fig:gaps}
\end{figure}

Theorem~\ref{thm:dual} provides us with a more physical representation of the BKM state ensemble; assuming that the Hamiltonian of the system can be described by the random model $P(H)$, then the state ensemble $P(\rho)$ will emerge given a thermalisation mechanism bringing any input to the Gibbs state associated with $H$. The question now becomes, what kind of system might be described by $P(H)$? While chaotic systems admit different random-matrix models depending on symmetries and constraints~\cite{d2016quantum}, their bulk local correlations are often universal. We can see that the BKM Hamiltonian ensemble falls into this universality class by the considering the distribution of neighbouring levels $E_\nu=E$ and $E_{\nu+1}=E+\delta$. Observe that the Vandermonde factors in~\eqref{eq:pE} scale quadratically with respect to the gap $\delta$,
\begin{align}
    \big|\big(e^{-E_{\nu+1}}-e^{-E_\nu}\big)(E_{\nu+1}-E_\nu)\big|=e^{-E}|\delta|^2+\mathcal{O}(\delta^3).
\end{align}
In addition, the potential $e^{-\frac{N^2}{2} F(H)}$ is strongly confining and exponentially suppresses the negative tail of eigenvalues. Hence the unfolded bulk spacing statistics is well defined and within the GUE universality class, with a  nearest-neighbour gap distribution approximated by the Wigner surmise \cite{mehta2004random}
\begin{align}\label{eq:wigner}
    p(s)=(32/\pi^2) s^2 e^{-4s^2/\pi}.
\end{align}
Figure~\ref{fig:gaps} shows a numerical simulation of $p(s)$ for our model which demonstrates the characteristic GUE level repulsion in excellent agreement with the surmise~\eqref{eq:wigner}. The algorithm to generate this histogram follows from Corollary~\ref{cor:alg}.

\begin{figure}[!t]
  \centering
  \includegraphics[width=\columnwidth]{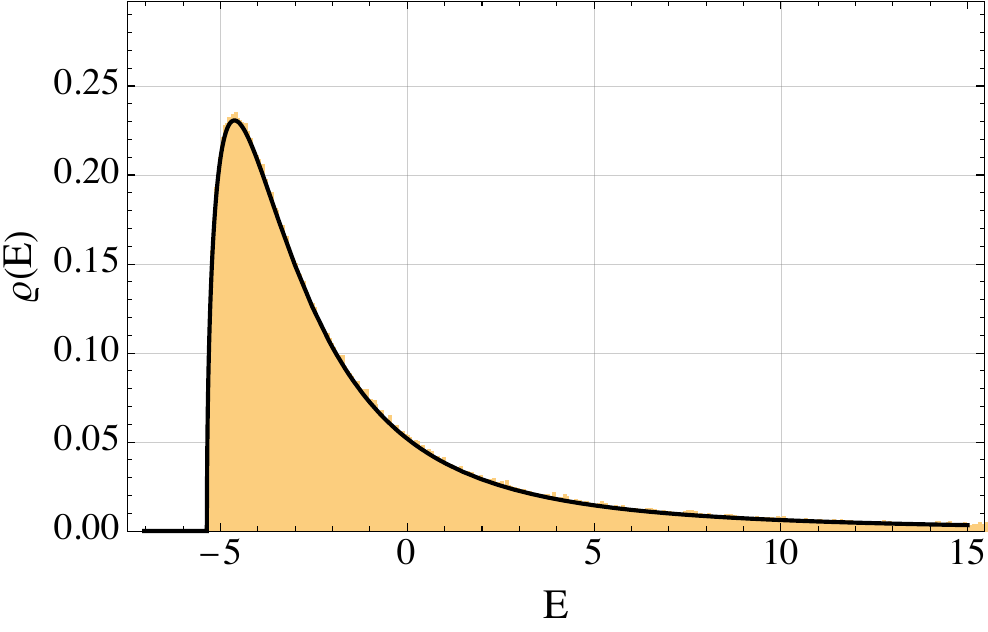}
  \caption{Energy level density of the BKM Hamiltonian ensemble in dimensionless units. The solid line is the analytic expression~\eqref{eq:rhoE}. Histogram was generated using the algorithm in Corollary~\ref{cor:alg}.  System size is $N=200$ and number of samples is $10^4$.  }
  \label{fig:density}
\end{figure}

On the other hand the global energy density significantly differs from the usual semicircle law of the GUE. To find its asymptotic expression we begin with the formula~\eqref{eq:lambert} we derived for eigenvalue density of the BKM ensemble. If $x=Nr$ denotes the rescaled eigenvalue of a BKM state $\rho$ distributed by $P(x)$, then according to  Theorem~\ref{thm:dual} we can relate this to the asymptotic energy density by introducing the variable $\epsilon=-\text{ln} \ x$ and transforming the density~\eqref{eq:lambert}:      
\begin{align}\label{eq:lambert2}
    P_\epsilon(\epsilon)=P(e^{-\epsilon})\bigg|\frac{d}{d\epsilon}e^{-\epsilon}\bigg|=-\Im \bigg[\frac{1}{\pi W(-2e^{\epsilon})}\bigg], 
\end{align}
where $\epsilon\in[-\text{ln} \ 2e,\infty]$. Given the traceless constraint $\sum_{j=1}^N \epsilon_j=0$ and assuming self averaging of the variable $\epsilon$,  the energy density is shifted by the mean $\bar{\epsilon}=\frac{1}{N}\sum_{j=1}^N \langle\epsilon_j \rangle$,
\begin{align}
\varrho(E)=P_\epsilon(\epsilon+\bar{\epsilon}), \ \ \ E\geq -\text{ln} \ 2e-\bar{\epsilon}
\end{align}
To estimate the mean $\bar{\epsilon}$ observe the upper tail 
\begin{align}
    P_\epsilon(\epsilon)\simeq \frac{1}{(\epsilon+\text{ln} \ 2)^2} \ \text{as} \ \epsilon\to\infty.
\end{align}
Furthermore, from our approximate exponential scaling of the minimum BKM eigenvalue~\eqref{eq:rankbound}, $x_{\min}\sim 2 e^{-N}$ we may fix a typical cutoff $\epsilon_\text{max}\sim N$ and estimate for large $N\gg 1$
\begin{align}
    \bar{\epsilon}\sim \frac{\int^{\epsilon_\text{max}}_{\epsilon_\text{min}}d\epsilon \ \epsilon P_\epsilon(\epsilon)}{\int^{\epsilon_\text{max}}_{\epsilon_\text{min}}d\epsilon \  P_\epsilon(\epsilon)}\simeq \text{ln} \ 2 \ \text{ln} \ N+\mathcal{O}(1)
\end{align}
In summary, the asymptotic energy density is 
\begin{align}\label{eq:rhoE}
    \varrho(E)=-\Im \bigg[\frac{1}{\pi  W(-2e^{E+\bar{\epsilon}})}\bigg], \ \ E\geq -\text{ln} \ 2e-\bar{\epsilon} 
\end{align}
where $\bar{\epsilon}=(\text{ln} \ 2) \text{ln} \ N+\mathcal{O}(1)$. In Figure~\ref{fig:density} we plot a numerical simulation of the energy density and fit the analytic prediction~\eqref{eq:rhoE}. 

The density is characterised by its heavy weighting and peak towards the lower energy states, qualitatively resembling the shape of the entanglement spectrum of a Haar random state \cite{geraedts2016many}. However a distinguishing feature of the BKM Hamiltonian density is its quadratic power law decay $\varrho(E)\sim 1/E^2$ for larger energies.

\section{Thermofield dynamics and chaos}\label{sec:7}

\

As an application of the matrix model we have introduced, we now turn to studying signatures of quantum chaos. The thermodynamic interpretation of the BKM ensemble naturally lends itself to model thermofield dynamics, which are particularly useful platforms for studying the impacts of quantum chaos on information scrambling and decoherence in closed quantum systems \cite{del2017scrambling,PhysRevB.103.064309,del2018decay}, as well as an important class of states in AdS/CFT \cite{hayden2007black,sekino2008fast}. A thermofield double (TFD) state represents a purification of a Gibbs state \cite{umezawa_matsumoto_tachiki_1982}, $\ket{TFD}\in\mathcal{H}\otimes \bar{\mathcal{H}}$ with
\begin{align}\label{eq:TFD}
    \ket{TFD}=\frac{1}{\sqrt{Z}}\sum_{n=1}^N e^{-\frac{1}{2}\beta E_n}\ket{n}_L\otimes \ket{n}_R.
\end{align}
Here $\ket{n}_{L/R}$ represents an energy eigenbasis in the left/right Hilbert space respectively and $Z=\tr{e^{-\beta H }}$ the partition function with respect to an inverse temperature scale $\beta$. By definition, a reduced state over one half of the Hilbert space is the Gibbs state
\begin{align}\label{eq:gibb}
    \rho=\text{Tr}_{\mathcal{\bar{H}}} \ket{TFD}\bra{TFD}=\frac{e^{-\beta H}}{\tr{e^{-\beta H}}}
\end{align}
If the adimensional matrix $\beta H$ is randomly distributed according to the BKM Hamiltonian ensemble $P(\beta H)$, then by Theorem~\ref{thm:dual} the states~\eqref{eq:gibb} are distributed by the BKM state ensemble~\eqref{eq:measure}. Thus a random TFD state~\eqref{eq:TFD} should be viewed as the set of random pure states that induce the mixed state ensemble $P(\rho)$, in analogy to how Haar distributed pure states give rise to the Hilbert-Schmidt~\eqref{eq:HS} and Bures-Hall ensembles~\eqref{eq:BH}. As a purification, the von-Neumann entropy of the reduced thermal state $\rho$ measures the entanglement of the TFD state, with its asymptotic value~\eqref{eq:asym} close to maximally entangled.

By setting the dimensionless timescale $\tau=t/\hbar\beta$ in units of the Planckian dissipation time we can represent the  evolution by the  unitary 
\begin{align}
    U_\tau=e^{-i \tau (H\otimes \mathbb{I}_R+ \mathbb{I}_L \otimes H)}, 
\end{align} 
with $H$ drawn from the BKM Hamiltonian ensemble $P(H)$. This random unitary dynamics gives us a model of chaotic TFD dynamics, and one way to probe this is through the statistics of the survival probability $F_\tau$. This measures the fidelity between the initial TFD state and its evolved state $\ket{\psi(\tau)}=U_\tau \ket{TFD}$, 
\begin{align}
    F_\tau=\big|\bra{TFD }U_\tau \ket{TFD}\big|^2=\bigg|\frac{\tr{e^{-(1+2i\tau)H}}}{\tr{e^{-H}}}\bigg|^2,
\end{align}
hence quantifying the spreading of information over time \cite{del2018decay}. The short time behaviour of the survival probability is determined by the average energy variance since
\begin{align}
    F_\tau=1-4 \tau^2\text{Var}_{\rho}(H)+\mathcal{O}(\tau^4).
\end{align}
For sufficiently large $N$ we can estimate the average variance under a standard self-averaging assumption for the measure, and compute it using the energy density
\begin{align}
    \nonumber\langle \text{Var}_{\rho}(H) \rangle&\approx \frac{d^2}{d\alpha^2}\text{ln}\int^\infty_{E_{\min}}dE \ \varrho(E)e^{-\alpha E}\bigg|_{\alpha=1}, \\
    \nonumber&=\frac{d^2}{d\alpha^2}\text{ln}\int^{2e}_{0}dx \ P(x) x^{\alpha+1}\bigg|_{\alpha=0}, \\
    &=\frac{9}{4}-\frac{\pi^2}{6}
\end{align}
where we used the Mellin transform identity~\eqref{eq:mellin}. The Zeno time $t_Z$ determines the timescale at which the onset of exponential decay in the survival probability begins \cite{del2017scrambling}, and in our model we estimate it as
\begin{align}
    t_Z=\frac{\hbar \beta}{2\sqrt{\langle \text{Var}_{\rho}(H) \rangle}}\approx 0.64 \  \hbar \beta.
\end{align}
and hence at the order of the Planckian dissipation time. In the long time limit the survival probability will plateau at a point fixed by the purity of the reduced thermal state,
\begin{align}
    \langle F_\infty \rangle=\langle\tr{\rho^2}\rangle\simeq \frac{8}{3N}, \ \ \ N\gg 1,
\end{align}
where we recall our earlier calculation of the purity in~\eqref{eq:purity}. 

\begin{figure}[!t]
  \centering
  \includegraphics[width=\columnwidth]{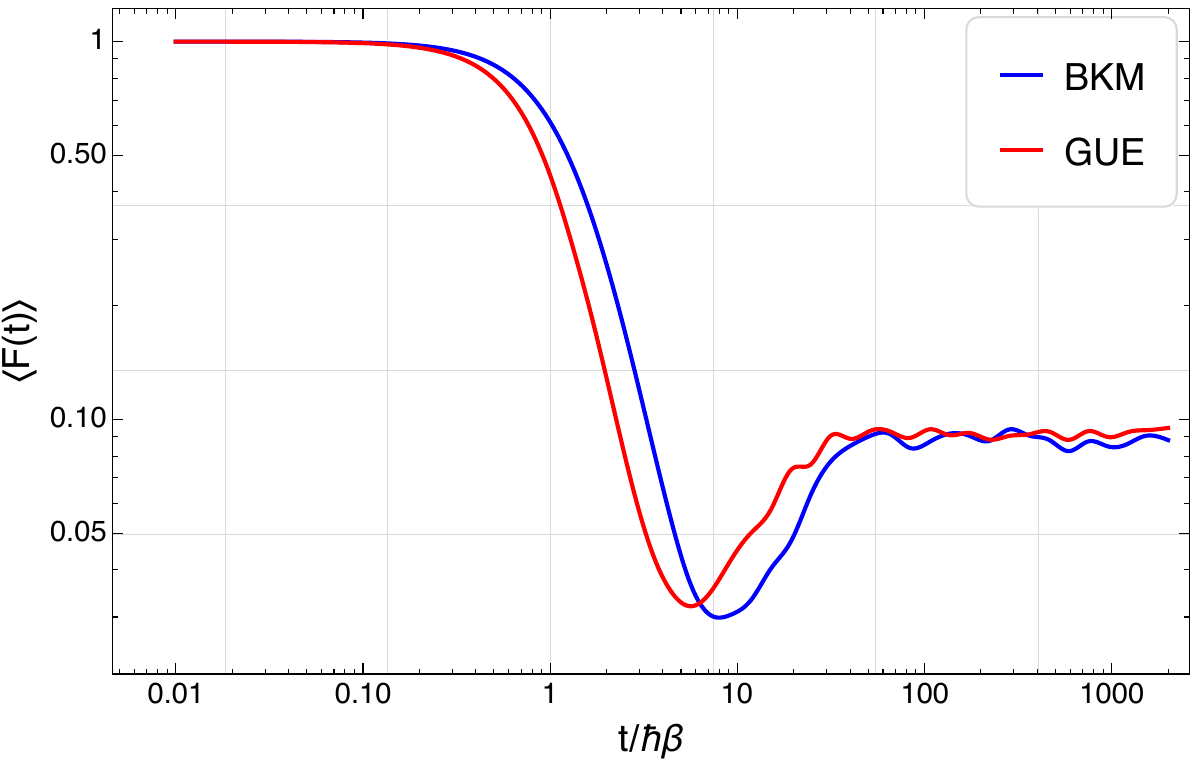}
  \caption{Average survival probability of an evolved TFD state as a function of dimensionless Planckian time $\tau=t/\hbar\beta$. Dynamics are simulated using the BKM Hamiltonian model (blue) and GUE ensemble (red). }
  \label{fig:TFD}
\end{figure}

In order to describe the full evolution of the survival probability we must also consider the impact of spectral correlations. Due to the Wigner level repulsion of the BKM model one expects a typical signature of chaos, namely a characteristic ramp and plateau in $\langle F_\tau \rangle$ arising from interference effects in the underlying spectrum. In Figure~\ref{fig:TFD} we have run a numerical simulation of the TFD dynamics for the BKM Hamiltonian model and plot the average survival probability, observing the expected ramp/plateau behaviour in $\langle F_\tau\rangle $. This qualitatively matches scrambling predicted by the GUE matrix model \cite{del2017scrambling,PhysRevB.103.064309,del2018decay}
\begin{align}
    P_{GUE}(H)\propto e^{-\frac{1}{4}N\tr{H^2}},
\end{align}
which is also compared in Figure~\ref{fig:TFD}. The simulation confirms that the BKM Hamiltonian model can capture some of the key signatures of quantum chaos.     

\section{Summary}\label{sec:8}

\

To summarise, we have presented an entropy-based method for sampling random density matrices according to the Bogoliubov-Kubo-Mori (BKM) metric $ds^2=-d^2 S(\rho)$ and investigated the statistical properties of the resulting distribution $P(\rho)$. This offers a new `\textit{third recipe}' for generating random quantum states, contrasting with previously known procedures based on the Hilbert-Schmidt (HS) and Bures-Hall (BH) ensembles. Our results further establish a novel connection between the Muttalib-Borodin random matrix ensemble and quantum information theory. Our key analytic calculation was the asymptotic eigenvalue density~\eqref{eq:lambert} and its associated entropy. A distinguishing feature of the BKM ensemble is its relatively smaller entropy~\eqref{eq:ineq} and exponentially decreasing minimum eigenvalue~\eqref{eq:rankbound}. We have left general finite-sized calculations of the entropy statistics and eigenvalue density as open problems.

With regards to quantum information theory, one application of the BKM state ensemble is as an uninformative prior for Bayesian state inference and tomography tasks. As has been noted in \cite{granade2016practical}, fiducial measures such as the HS and BH ensembles are not always appropriate priors for describing experiments that feature low-rank, high purity states such as in entanglement distillation. While this may be circumvented by fine-tuning priors to specific experiments \cite{mai2017pseudo,lohani2021improving},  it is also desirable to have a default prior that can be used across varying experimental settings over time. The effective rank deficiency of the BKM ensemble means it could be used as an uninformative prior in such situations. Moreover, it would be interesting to combine it with adaptive Bayesian methods \cite{lohani2023demonstration} and test its relative performance in state tomography experiments.

Alongside the BKM states we also derived a dual Hamiltonian ensemble $P(H)$ using the thermodynamic length  metric on the space of Gibbs states. This Legendre duality between the two ensembles lead to a thermodynamic interpretation of the BKM states (Theorem~\ref{thm:dual}); they are equivalent to an ensembles of  Gibbs states distributed according to the random Hamiltonian model $P(H)$. In essence this Hamiltonian ensemble can be viewed as a quantum analogue of Ruppeiner's classical fluctuation theory \cite{ruppeiner1995riemannian}, which posits a covariant model of macroscopic equilibrium fluctuations measured in terms of thermodynamic length \cite{crooks2007measuring}. We demonstrated that our model recovers GUE level repulsion in the thermodynamic limit, suggesting that the BKM ensemble can serve as a thermodynamically motivated model of quantum chaos. A future research direction could be to understand how such a model can emerge dynamically via Dyson Brownian motion in the context of noisy quantum evolution \cite{gerbino2024dyson}. One can also interpret this ensemble as a random modular Hamiltonian model with its level statistics quantifying the entanglement spectrum of randomly sampled thermofield double states. This physical interpretation suggests that the BKM ensemble should be well suited to describing models of semi-classical gravity \cite{cotler2017black,altland2021late,de2024principle}. In fact, the BKM entropy-based metric itself has been shown to play a fundamental role in the AdS/CFT correspondence since its gravitational dual is canonical energy \cite{lashkari2016canonical}.

Due to the fundamental relation between the BKM metric and statistical mechanics it would be promising to explore the connection between the BKM random state ensemble and notions such as typicality \cite{gogolin2016equilibration} and deep thermalisation \cite{mark2024maximum}.  Lastly, the present paper has only considered finite-dimensional quantum systems, and it would be beneficial to extend the BKM ensemble to certain infinite-dimensional systems such as Gaussian states \cite{miller2025curvature}.

\textit{Acknowledgments:} H.M. acknowledges funding from a Royal Society Research Fellowship (URF/R1/231394). I also thank Karen Hovhannisyan  for helpful discussions.

\bibliographystyle{apsrev4-1}
\bibliography{mybib2.bib}

\appendix

\widetext

\section{Normalisation constant}\label{app:A}

\

To find the normalisation constant for the distribution of eigenvalues~\eqref{eq:pdf} we first need to compute the following integral
\begin{align}
    \nonumber \tilde{R}(s)&:=\int_0^\infty dr_1 dr_2\cdots dr_N \ \prod_{\nu<\mu} (r_\mu-r_\nu) \text{ln}[r_\mu/r_\nu]\prod^N_{k=1}r_k^{-1/2}e^{-s r_k}, \\
    \nonumber&=\int_0^\infty d(sr_1) d(sr_2)\cdots d(sr_N) \ s^{-N^2/2}\prod_{\nu<\mu} (sr_\mu-sr_\nu) \text{ln}[sr_\mu/sr_\nu]\prod^N_{k=1}(sr_k)^{-1/2}e^{- s r_k}, \\
    &=s^{-N^2/2}\int_0^\infty dr_1 dr_2\cdots dr_N \ \prod_{\nu<\mu} (r_\mu-r_\nu) \text{ln}[r_\mu/r_\nu]\prod^N_{k=1}r_k^{-1/2}e^{-  r_k}
\end{align}
Recall now the well-known formula for the Vandermonde determinant \cite{mehta2004random},
\begin{align}
    \prod_{\nu<\mu}(a_\mu-a_\nu)=\text{det}[a_\nu^{\mu-1}]_{\nu,\mu=1,\cdots, N}
\end{align}
so the integral becomes
\begin{align}\label{eq:Rs}
    \tilde{R}(s)=s^{-N^2/2}\int_0^\infty dr_1 dr_2\cdots dr_N \ \text{det}[r^{\mu-3/2}_\nu e^{- r_\nu}]_{\nu,\mu=1,\cdots ,N}\text{det}[\text{ln}^{\mu-1}(r_\nu)]_{\nu,\mu=1,\cdots ,N},
\end{align}
Now we invoke Andr\'eief’s integral formula \cite{forrester2019meet}: let $\{f_\mu(x)\}_{\mu=1,..,N}$ and $\{g_\mu(x)\}_{\mu=1,..,N}$ be two sequences of integrable functions. Then the following identity holds:
\begin{align}
    \frac{1}{N!}\int_0^\infty dx_1 dx_2...dx_N \ \text{det}[f_\mu(x_\nu)]_{\nu,\mu=1,\cdots, N} \ \text{det}[g_\mu(x_\nu)]_{\nu,\mu=1,\cdots, N}=\text{det}[\braket{f_\mu|g_\nu}]_{\nu,\mu=1,\cdots, N},
\end{align}
where 
\begin{align}
    \braket{f|g}=\int_0^\infty dx \ f(x)g(x)
\end{align}
Therefore we need the integral
\begin{align}
    \int^\infty_0 dx \ x^{\mu-3/2}e^{- x} \text{ln}^{\nu-1}(x)=\Gamma^{(\nu-1)}(\mu-1/2),
\end{align}
where $\Gamma^{(n)}(z)$ is the $n$'th derivative of the Gamma function. Combining this with~\eqref{eq:Rs} one gets
\begin{align}
    \tilde{R}(s)=N! \ s^{-N^2/2}\text{det}\big[\Gamma^{(\nu-1)}(\mu-1/2)\big]_{\nu,\mu=1,\cdots, N}, 
\end{align}
Now set $z_\mu=\mu-1/2$ and $D_N=\text{det}\big[\Gamma^{(\nu-1)}(z_\mu)\big]_{\nu,\mu=1,\cdots, N}$. For each column vector $c_\mu=[\Gamma(z_\mu),\Gamma^{(1)}(z_\mu),\cdots]^T$ we perform a determinant-persevering operation $c_\mu\mapsto c_\mu-z_{\mu-1}c_{\mu-1}$ for $\mu\geq 2$. It follows from the identity $\Gamma(z+1)=z\Gamma(z)$ that 
\begin{align}
    \Gamma^{(m)}(z+1)=z\Gamma^{(m)}(z)+m\Gamma^{(m-1)}(z),
\end{align}
and so we may write the determinant $D_N=\text{det}[M_{\nu\mu}]_{\nu,\mu=1,\cdots N}$ where
\begin{align}
M=\left[
\begin{array}{ccccc}
\Gamma(z_1) & 0 & 0 & \cdots & 0 \\[4pt]
\Gamma^{(1)}(z_1) & 1\,\Gamma(z_1) & 1\,\Gamma(z_2) & \cdots & 1\,\Gamma(z_{N-1}) \\[4pt]
\Gamma^{(2)}(z_1) & 2\,\Gamma^{(1)}(z_1) & 2\,\Gamma^{(1)}(z_2) & \cdots & 2\,\Gamma^{(1)}(z_{N-1}) \\[4pt]
\vdots & \vdots & \vdots & \ddots & \vdots \\[4pt]
\Gamma^{(N-1)}(z_1) & (N-1)\,\Gamma^{(N-2)}(z_1) & (N-1)\,\Gamma^{(N-2)}(z_2) & \cdots & (N-1)\,\Gamma^{(N-2)}(z_{N-1})
\end{array}
\right]
\end{align}
Expanding the determinant along the first row gives 
\begin{align}
    \nonumber D_N&=\Gamma(1/2)\text{det}[(\nu-1)\Gamma^{(\nu-2)}(z_{\mu-1})]_{\nu,\mu=2,\cdots N}, \\
    \nonumber&=\Gamma(1/2)(N-1)! \ \text{det}[\Gamma^{(\nu-2)}(z_{\mu-1})]_{\nu,\mu=2,\cdots N}, \\
    &=\sqrt{\pi}(N-1)! D_{N-1}
\end{align}
By induction,
\begin{align}
    D_N=\pi^{N/2} \prod^{N-1}_{k=1} k!,
\end{align}
and therefore 
\begin{align}
    \tilde{R}(s)=\pi^{N/2} s^{-N^2/2} \prod^{N}_{k=1} k!
\end{align}
Finally the normalisation constant is related to $\tilde{R}(s)$ via an inverse Laplace transform,
\begin{align}
    \nonumber C^{-1}_N&=\int_0^\infty dr_1 dr_2\cdots dr_N \ \delta\bigg(1-\sum^{N}_{k=1} r_k\bigg) \prod_{\nu<\mu} (r_\mu-r_\nu) \text{ln}[r_\mu/r_\nu]\prod^N_{k=1}r_k^{-1/2}, \\
    \nonumber&=\mathcal{L}_s\bigg[\int_0^\infty dr_1 dr_2\cdots dr_N \ \prod_{\nu<\mu} (r_\mu-r_\nu) \text{ln}[r_\mu/r_\nu]\prod^N_{k=1}r_k^{-1/2}e^{-s r_k}\bigg](t)\bigg|_{t=1}, \\
    \nonumber&=\mathcal{L}_s\big[\tilde{R}(s)\big](t)\bigg|_{t=1}, \\
    &=\frac{\pi^{N/2}  \prod^{N}_{k=1} k!}{\Gamma(N^2/2)},
\end{align}
completing the derivation.

\section{Generating BKM states}\label{app:B}

\

In this section we will prove that the random state generation described by the algorithm in Theorem~\ref{thm:algo} from the main text correctly samples density matrices according to the BKM ensemble~\eqref{eq:measure}. Let $X$ be a lower triangular complex matrix sampled according to the prescription in Step A of the main text and let $\{s_1,\cdots . s_N\}$ denote its squared singular values. We may then construct a hermitian matrix  
\begin{align}
    XX^\dagger=V D(\vec{s}) V^\dagger, \ \ \ D=\text{diag}[s_1,\cdots, s_N]
\end{align}
with $V$ some arbitrary unitary matrix. By Theorem 2 of \cite{cheliotis2018triangular}, the singular values are distributed according to the Muttalib-Borodin ensemble
\begin{align}
    P_{MB}(\vec{s})=\text{const}.\prod_{k=1}^{N} \frac{e^{-s_k}}{\sqrt{s_k}}\prod_{\nu<\mu} (s_\mu-s_\nu)\text{ln}  \bigg(\frac{s_\mu}{s_\nu}\bigg)
\end{align}
As stated in the main text, this corresponds to the parameter choice $\alpha=-1/2$ and $\theta=0$ of the more general ensemble $P_{\theta,\alpha}(\vec{r})$ defined in~\eqref{eq:MB}. If we consider the full process Step A to C, we form the density matrix 
\begin{align}
    \rho=\frac{U XX^\dagger U^\dagger}{\Tr{U XX^\dagger U^\dagger}},
\end{align}
where $U\in U(N)$ is a Haar unitary. Let $\tilde{P}(\rho)$ be the probability density of the resulting states. Our goal is to show it is proportional to the BKM ensemble~\eqref{eq:measure}. By construction, this distribution takes the form
\begin{align}\label{eq:Prho}
    \nonumber \tilde{P}(\rho)&\propto\int_{U(N)}d\mu_{U}(U)\int_{\mathbb{R}_+^N} ds_1\cdots s_N \ P_{MB}(\vec{s})\delta\bigg[\rho-\frac{UV D(\vec{s}) V^\dagger U^\dagger}{\Tr{UV D(\vec{s}) V^\dagger U^\dagger}}\bigg], \\
    \nonumber&=\int_{U(N)}d\mu_{U}(U)\int_{\mathbb{R}_+^N} ds_1\cdots s_N \ P_{MB}(\vec{s})\delta\bigg[\rho-\frac{U D(\vec{s})  U^\dagger}{\Tr{ D(\vec{s}) }}\bigg], \\
    &\propto\int_{U(N)}d\mu_{U}(U)\int_{\mathbb{R}_+^N} ds_1\cdots s_N \ e^{-\sum^{N}_{k=1}s_k }\prod_{k=1}^{N} \frac{1}{\sqrt{s_k}}\prod_{\nu<\mu} (s_\mu-s_\nu)\text{ln}  \bigg(\frac{s_\mu}{s_\nu}\bigg)\delta\bigg[\rho-\frac{U D(\vec{s})  U^\dagger}{\Tr{ D(\vec{s}) }}\bigg], 
\end{align}
where we used the unitary invariance of the Haar measure to remove dependence on $V$. Now introduce a delta function via a dummy variable $t$, 
\begin{align}
    \tilde{P}(\rho)
    =\int_{U(N)}d\mu_{U}(U)\int_{0}^{\infty} dt\int_{\mathbb{R}_+^N} ds_1\cdots ds_N \ 
    e^{-t}\prod_{k=1}^{N} s_k^{-1/2}\prod_{\nu<\mu} (s_\mu-s_\nu)\ln\!\bigg(\frac{s_\mu}{s_\nu}\bigg)
    \ \delta\!\bigg[\rho-\frac{U D(\vec{s})U^\dagger}{t}\bigg]\ \delta\!\big(\Tr{D(\vec{s})}-t\big).
\end{align}
Setting \(s_k=t r_k\) and transforming variables we obtain
\begin{align}
    \nonumber\tilde{P}(\rho)
    =\int_{U(N)}d\mu_{U}(U)\int_{0}^{\infty} dt\int_{\mathbb{R}_+^N} dr_1\cdots dr_N \ 
    e^{-t}\,t^{\,\frac{N^2}{2}-1}\,
    \Big(\prod_{k=1}^{N} r_k^{-1/2}\Big)
    \Big(\prod_{\nu<\mu} (r_\mu-r_\nu)\ln\!\bigg(\frac{r_\mu}{r_\nu}\bigg)\Big)\,
    \delta\!\big[\rho-U D(\vec{r})U^\dagger\big]\ \delta\!\bigg(\sum_{k=1}^N r_k-1\bigg).
\end{align}
Finally, performing the \(t\)-integral gives
\begin{align}
    \nonumber\tilde{P}(\rho)
    &=\Gamma\!\left(\frac{N^2}{2}\right)\int_{U(N)}d\mu_{U}(U)\int_{\mathbb{R}_+^N} dr_1\cdots dr_N \ 
    \Big(\prod_{k=1}^{N} r_k^{-1/2}\Big)
    \Big(\prod_{\nu<\mu} (r_\mu-r_\nu)\ln\!\bigg(\frac{r_\mu}{r_\nu}\bigg)\Big)\,
    \delta\!\big[\rho-U D(\vec{r})U^\dagger\big]\ \delta\!\bigg(\sum_{k=1}^N r_k-1\bigg) \\
    &\propto \int_{U(N)}d\mu_{U}(U)\int_{\mathbb{R}_+^N} dr_1\cdots dr_N \ P(\vec{r})\,
    \delta\!\big[\rho-U D(\vec{r})U^\dagger\big],
\end{align}
which shows that the sampling procedure reproduces the BKM ensemble~\eqref{eq:measure} (up to normalisation). This concludes the proof.

\section{Volume form for thermodynamic length}\label{app:C}

\

Starting with the diagonal representation of the Hamiltonian, $H=\sum_k E_k \ket{k}\bra{k}$, the thermodynamic length is \cite{balian2014entropy}
\begin{align}\label{eq:Hmetric}
    ds^2=e^{-F(H)}\sum_{\nu,\mu} \bigg[\frac{e^{-E_\nu}-e^{-E_\mu}}{E_\mu-E_\nu}\bigg]|dH_{\nu\mu}|^2-e^{-2F(H)}\bigg(\sum_k e^{-E_k}dH_{kk}\bigg)^2
\end{align}
Setting $D=\text{diag}[E_1,\cdots E_N]$, we parameterise the Hamiltonian in diagonal form $U D U^\dagger$. Variations in $H$ can then be expressed in terms of an anti-hermitian matrix $\omega=-\omega^\dagger$, namely
\begin{align}
    dH=U(dD+[\omega, D]) U^\dagger,
\end{align}
where $dD=\text{diag}[dE_1,\cdots dE_N]$. We can then decompose this in the eigenbasis of $H$ with 
\begin{align}
    \nonumber &dH_{ii}=dE_i, \\
    \nonumber & dH_{ij}=(E_j-E_i)\omega_{ij}, \ \ i\neq j
\end{align}
Plugging this into~\eqref{eq:Hmetric} diagonalises the metric $ds^2=ds^2_{\text{diag}}+ds^2_{\text{off-d}}$ where
\begin{align}
    \nonumber ds^2_{\text{diag}}&=\sum_{\nu,\mu}(p_\nu \delta_{\nu\mu}-p_\nu p_\mu)dE_\nu dE_\mu, \\
    \nonumber ds^2_{\text{off-d}}&=2\sum_{\nu<\mu} (p_\mu-p_\nu)(E_\nu-E_\mu)\big[\big(\Re \omega_{\nu\mu}\big)^2+\big(\Im \omega_{\nu\mu}\big)^2\big]
\end{align}
and we set $p_i=e^{-(E_i+F(H))}$ for shorthand. Define the matrix 
\begin{align}
    G_{ij}=p_i \delta_{ij}-p_i p_j,
\end{align}
which clearly has a null eigenvalue with eigenvector $(1,\cdots, 1)^T$. We therefore need the pseudo-determinant $\text{det} [G^+]$ on the subspace orthogonal to $(1,\cdots, 1)^T$. To do so introduce a regularisation $G_\epsilon= G+\epsilon \mathbb{I}_N$ and look at the small $\epsilon$ behaviour $\text{det}[G_\epsilon]=\epsilon \text{det} [G^+]+\mathcal{O}(\epsilon^2)$ which implies
\begin{align}
    \text{det} [G^+]=\lim_{\epsilon\to 0} \frac{\text{det}[G_\epsilon]}{\epsilon}.
\end{align}
We define the invertible matrix $D_\epsilon=\text{diag}[(p_1+\epsilon),\cdots,(p_N+\epsilon)]$ and vector $\vec{p}=(p_1,\cdots,p_N)$. By the matrix determinant lemma,
\begin{align}
    \text{det}[G_\epsilon]=\text{det}\big[(p_i+\epsilon) \delta_{ij}-p_i p_j\big]_{i,j=1}^N=\text{det}[(p_i+\epsilon)]_{i=1}^N\bigg(1-\vec{p}^T D^{-1}_\epsilon \vec{p} \bigg).
\end{align}
It follows from the normalisation $\sum_i p_i=1$ that
\begin{align}
    \vec{p}^T D^{-1}_\epsilon \vec{p}=\sum_i \frac{p^2_i}{p_i+\epsilon}=1-\epsilon\sum_i \frac{p_i}{p_i+\epsilon}+\mathcal{O}(\epsilon)
\end{align}
and hence 
\begin{align}
    \text{det}[G_\epsilon]=\epsilon\bigg(\prod_i (p_i+\epsilon)\bigg)\sum_i \frac{p_i}{p_i+\epsilon}=\epsilon \sum_i p_i \prod_{j\neq i}(p_j+\epsilon)
\end{align}
Finally we compute 
\begin{align}
    \text{det} [G^+]=\lim_{\epsilon\to 0} \sum_i p_i \prod_{j\neq i}(p_j+\epsilon)=\sum_i p_i\prod_{i\neq j} p_j=N \prod^N_{i=1}p_i.
\end{align}
Returning to the diagonalised metric $ds^2=ds^2_{\text{diag}}+ds^2_{\text{off-d}}$ we can now compute the volume form,
\begin{align}
    \nonumber dV(H)&=2^{N(N-1)/2} \sqrt{\text{det}[G^+]}\bigg(\prod_{\nu<\mu} (p_\mu-p_\nu)(E_\nu-E_\mu)\bigg) d^N \vec{E}\times d\mu_{U}(U), \\
    \nonumber&=\sqrt{N} 2^{N(N-1)/2} \prod^N_{i=1}\sqrt{p_i}\prod_{\nu<\mu} (p_\mu-p_\nu)(E_\nu-E_\mu) d^N \vec{E}\times d\mu_{U}(U), \\
    \nonumber&=\sqrt{N} 2^{N(N-1)/2}\ e^{-\frac{N^2}{2}F(H)}\prod^N_{i=1}e^{-\frac{1}{2}E_i}\prod_{\nu<\mu} (e^{-E_\mu}-e^{-E_\nu})(E_\nu-E_\mu) d^N \vec{E}\times d\mu_{U}(U), \\
    &=\sqrt{N} e^{-\frac{N^2}{2}F(H)}\prod_{\nu<\mu} (2 e^{-E_\mu}-2e^{-E_\nu})(E_\nu-E_\mu) d^N \vec{E}\times d\mu_{U}(U),
\end{align}
where we used $\tr{H}=0$ in the final line. This can be rewritten in terms of the hermitian measure 
\begin{align}
    \mathcal{D}H=\Delta^2(H)d^N \vec{E}\times d\mu_{U}(U)
\end{align}
and we can compose a probability density for the Hamiltonian,
\begin{align}
P(H)\mathcal{D}H=\delta(\tr{H})\frac{dV(H)}{\text{Vol}(N)}= \frac{\sqrt{N}e^{-\frac{N^2}{2} F(H)}}{\text{Vol}(N)}\delta(\tr{H}) \bigg|\frac{\Delta(2 e^{-H})}{\Delta(H)}\bigg|\mathcal{D}H.
\end{align}
which completes the derivation of~\eqref{eq:pH}. Note that the volume $\text{Vol}(N)$ is the same as~\eqref{eq:volume} computed for the BKM ensemble on $\mathcal{S}(\mathcal{H})$; this follows from Theorem 1 since  volume is preserved under the isometry relating $dV(H)$ and $dV(\rho)$.

\end{document}